\documentclass[nonacm, manuscript]{acmart} 

\usepackage{url}
\usepackage{multicol}
\usepackage{multirow}
\usepackage{graphicx}  
\usepackage{wrapfig}
\usepackage{mdframed}
\usepackage{xcolor}
\usepackage{comment}
\usepackage{ragged2e}
\usepackage{xspace}
\usepackage{adjustbox}
\usepackage{listings}
\usepackage{tikz}  
\usetikzlibrary{shapes.geometric, arrows}  
  
\tikzstyle{block} = [rectangle, rounded corners, minimum width=2cm, minimum height=0.8cm, text centered, draw=black, align=center] 
\tikzstyle{arrow} = [thick,->,>=stealth]  

\newcommand{\NoAI}{NoAI\xspace}

\newmdenv[
    backgroundcolor=gray!10,
    hidealllines=true,
    leftline=true,
    linecolor=gray!75, 
    linewidth=3pt,
    innerleftmargin=15pt,
    innerrightmargin=10pt,
    innertopmargin=10pt,
    innerbottommargin=10pt
]{grayquote}

\newmdenv[
    topline=false,
    rightline=false,
    bottomline=false,
    leftline=false,
    innertopmargin=0pt,
    innerbottommargin=5pt,
    innerleftmargin=0pt,
    innerrightmargin=0pt,
]{finding}

\definecolor{darkheading}{gray}{0.80}
\definecolor{lightbody}{gray}{0.95}
  
\newcommand{\IITK}{IIT Kanpur}  
\newcommand{\amount}{\textsc{INR 150}}  
  
\begin{document}


\title{To Google or To ChatGPT? A Comparison of CS2 Students' Information Gathering Approaches and Outcomes}






\author{Aayush Kumar}
\affiliation{%
  \institution{Indian Institute of Technology Kanpur}
  \country{India}}
\affiliation{%
  \institution{Microsoft}
  \city{Bengaluru}
  \country{India}}
\email{t-aaykumar@microsoft.com}

\author{Daniel Prol}
\affiliation{%
  \institution{Universidad Internacional de la Rioja}
  \country{Spain}}
\email{daniel.prol403@comunidadunir.net}

\author{Amin Alipour}
\affiliation{%
  \institution{University of Houston	}
  \country{United States}}
\email{maalipou@central.uh.edu	}

\author{Sruti Srinivasa Ragavan	}
\affiliation{%
  \institution{Indian Institute of Technology Kanpur	}
  \country{India}}
\email{srutis@cse.iitk.ac.in}













\begin{abstract}

LLMs such as ChatGPT have been widely adopted by students in higher education as tools for learning programming and related concepts. However, it remains unclear how effective students are and what strategies students use while learning with LLMs. Since the majority of students' experiences in online self-learning have come through using search engines such as Google, evaluating AI tools in this context can help us address these gaps. In this mixed-methods research, we conducted an exploratory within-subjects study to understand how CS2 students learn programming concepts using both LLMs as well as traditional online methods such as educational websites and videos to examine how students approach learning within and across both scenarios. We discovered that students found it easier to learn a more difficult concept using traditional methods than using ChatGPT. We also found that students ask fewer follow-ups and use more keyword-based queries for search engines while their prompts to LLMs tend to explicitly ask for information.

\end{abstract}


\settopmatter{printacmref=false}
\setcopyright{none}
\renewcommand\footnotetextcopyrightpermission[1]{}
\pagestyle{plain}
\fancyfoot{}
\maketitle


\section{Introduction}

With the widespread adoption of digital education \cite{weller2022rise},
understanding the tools and processes behind self-directed (SDL) and self-regulated learning (SRL) has become increasingly important. Online educational resources tend to be the primary tools for self-directed learners \cite{Mello_2016}, especially in the case of learning programming \cite{Mccartney2010ComputingSL}. The advancement of generative artificial intelligence (AI) has introduced a variety of online tools and methods to support SDL \cite{ai4sdl}. In particular, Large Language Models (LLMs) such as ChatGPT have gained popularity for their ability to assist learners through interactive and personalized guidance \cite{amoozadeh2024interaction}. Unlike searching through the web, which often requires complex navigation between multiple sources \cite{sellen2002knowledge}, these models respond directly to user queries in real-time. They can generate examples as well as high-quality explanations \cite{leinonen2023comparing, denny2023can}, making them a promising resource for learning programming \cite{kazemitabaar2023novices}.
However, the path to their adoption for self-directed programming learning is replete with challenges: these tools sometimes generate inaccurate or misleading information, lack the ability to understand nuanced learner needs, and potentially foster over-reliance on AI instead of critical thinking and independent problem solving ~\cite{ASarkar2022WhatIsItLike}. Thus, research on whether or not to adopt LLMs in learning contexts is divided: some studies such as \cite{azaiz2023ai,prather2023robots, sun2024promptlearning, amoozadeh2024interaction, kazemitabaar2023novices, guo2023six} indicate their promise, while others caution against their use~\cite{prather2023weird, prather2024wideninggapbenefitsharms}.

As the number of online resources for learning programming increases, so does the complexity of information problems for self-directed learners \cite{manyresource}.
However, there is a paucity of empirical studies comparing information gathering approaches with LLMs to traditional web-based learning (e.g., using video tutorials, documentation, and websites such as Stack Overflow) -- both in terms of quantitative outcomes and qualitative behaviors \cite{10.1145/3632620.3671112}. As a result, making informed decisions about learning choices becomes difficult for both students and educators alike.

 
In this paper, we begin to address this gap by comparing how learners gather information about programming concepts with AI vs. traditional web (\NoAI) resources such as Google search and YouTube videos. Specifically, we conducted a within-subject lab study at \IITK, a large engineering university in India, involving 32 participants (CS2 students) learning new topics in programming concepts with LLMs and traditional web resources. We investigated the differences in the strategies learners adopt in both these information-seeking setups, along with their underlying reasons, and the effectiveness of these strategies in terms of assessments. Our results suggest that the participants performed better on the post-study evaluations in the \NoAI condition. We also observed a statistically significant difference between the types of prompts in AI and queries under \NoAI conditions. 

Our key contributions are: 1) a comparison of students' information-seeking behavior in learning tasks for two programming concepts via a lab study, and 2) an assessment and comparison of the effectiveness of students' information-seeking strategies through quizzes and debugging tasks when using AI and \NoAI.

\section{Related Work}

\paragraph{Student-AI Interaction in Programming Education}

A review of research on student interactions with AI tools in programming education reveals both opportunities and challenges. 
Prather et al.~\cite{prather2023robots, prather2023weird} reviewed the use of LLMs in computing education, highlighting their potential to enhance learning while warning against overreliance. AI-based feedback mechanisms show promise in supporting students~\cite{azaiz2023ai,macneil2023experiences,10.1145/3613904.3642773,roest2023nextstephintgenerationintroductory}, though balancing automation with pedagogy remains a challenge. Kazemitabaar et al.~\cite{kazemitabaar2023novices} categorized novice programming strategies and found that hybrid approaches (manual coding + AI assistance) yield better results. Amoozadeh et al.~\cite{amoozadeh2024trust} examined trust dynamics, showing how perception influences AI adoption. Researchers agree that integrating AI in a structured and deliberate manner can enhance self-regulation and programming skills for CS students in higher education \cite{prather2023robots}. This deliberate integration calls for an understanding of how students gather information when learning with LLMs and the scenarios in which their strategies are most effective.

\paragraph{ChatGPT as a Learning Tool}
The conversational interface of ChatGPT can enhance programming education by enabling personalized learning, as shown by Sun et al.~\cite{sun2024promptlearning} and Amoozadeh et al.~\cite{amoozadeh2024interaction}. Studies by Kazemitabaar et al.~\cite{kazemitabaar2023novices} and Guo et al.~\cite{guo2023six} highlight its ability to aid debugging and concept clarification. 
Prior research has indicated that \textit{prompt strategies} play a key role in determining the effectiveness of ChatGPT. Sun et al.~\cite{sun2024prompt} reported that structured prompts improve outcomes, while Nguyen et al.~\cite{nguyen2024beginning} identified challenges in prompt formulation and solution evaluation.  Lee et al.~\cite{lee2025impact} found that generative AI shifts effort from searching to verifying, requiring stronger prompt and validation skills. This information-seeking gap is reflected in Prather et al.'s ~\cite{prather2024widening} findings that while AI accelerates learning for some students, it may hinder others. Skripchuk et al. \cite{10.1145/3632620.3671112} investigated student intentions behind using the web or LLMs to learn programming, highlighting that comparisons of information-seeking strategies between LLM-based and traditional self-study methods remain scarce. A study by Qureshi et al.~\cite{Qureshi_2023} found students using ChatGPT outperformed those relying solely on offline materials for a programming assessment. Our study extends this line of work by comparing AI-assisted and web-based self-study in programming education.

\paragraph{Online Information-Seeking Strategies in Programming}  
Prior research on information-seeking behavior has identified key patterns in how developers learn programming online. Sellen et al.~\cite{sellen2002knowledge} classified information gathering tasks as a core aspect of knowledge work for developers, accounting for 35\% of web activities. These tasks often require navigating multiple sources to gain a holistic understanding. 
Ko et al.~\cite{ko2007information} mapped common developer information needs, while Upadhyay et al.~\cite{upadhyay2020comparing} explored how interface modality affects search behavior. Duala-Ekoko and Robillard~\cite{duala2014information} identified three primary information-seeking strategies in programming learning: web-based (seeking code online), documentation-based (official manuals), and hybrid (combining both). Prior work has also described how developers manage information-seeking tasks through strategies such as opportunistic programming ~\cite{brandt2009twostudies}, in which developers adapt online examples to their needs. 
With AI emerging as an important online resource for programmers \cite{stackoverflow2024}, there is a need to understand how these strategies compare to those of gathering information with AI tools.

\paragraph{AI and Web-Based Resources in Self-Directed Learning}  
Self-directed learning (SDL) is an essential part of programming education, fostering problem-solving skills and adaptability ~\cite{wirth2024self}. Computing students engage in SDL through personal projects, social influences, and skill competitiveness ~\cite{mccartney2016computing}. AI-driven tools are reshaping SDL by providing personalized instruction and automated feedback ~\cite{sun2024investigating} ~\cite{mzwri2025ai}. Younas et al.~\cite{younas2025systematic} review AI-based SDL approaches, highlighting key tools such as adaptive tutoring and real-time feedback. Structured AI interactions enhance engagement, as seen in ChatGPT-based learning environments~\cite{sun2024investigating}. While prior research indicates the promise of AI for SDL,  it is not without its obstacles. Wirth et al.~\cite{wirth2024self} suggest that the effectiveness of AI tools depends on structured pedagogy, ensuring students engage critically rather than relying passively on AI. Studies indicate that students with prior knowledge and digital literacy benefit the most from AI-based SDL, highlighting the need for structured scaffolding~\cite{younas2025systematic}. Examining learners' information-seeking strategies with AI in the context of traditional methods can facilitate this structured adoption of AI by providing insights into how different tools benefits students in a self-directed learning environment.






\section{Research Questions}
We aim to address the following research questions:
\begin{enumerate}
    \item What, if any, are the differences in information-seeking strategies between the web and AI treatments?
    \item How effective are the information-seeking strategies in the web and AI treatments in terms of conceptual understanding and practical application? 
\end{enumerate}

\section{Methodology}
To answer the above questions, we conducted a user study with 32 participants, comparing their information-seeking strategies when learning programming concepts with and without AI. 
The study was approved by the Institutional Ethics Committee at \IITK.

\subsection{Study Design}
Our study used an in-vitro, within-subject design. Each participant learned two different programming concepts, namely currying ~\cite{CurryingInJavaScript} and immediately invoked function expressions (IIFE) ~\cite{IIFEInJavaScript} in JavaScript. They were instructed to learn one of these tasks using only ChatGPT (model version GPT 3.5 Turbo), and the other using traditional online resources (e.g web search, videos) without the use of any generative AI tools starting with a Google search page. 
We call these two conditions AI and \NoAI respectively. Participants in both conditions had access to the Visual Studio Code (VSCode) IDE to run code if they decided. The task assignment to conditions and task order were balanced.

\begin{table}[h]  
\centering  
\begin{tabular}{|c|c|c|}  
\hline  
 \textbf{Computer Science} & \textbf{Engineering} & \textbf{Sciences and Others} \\
 \hline  
 6 & 20 & 6 \\
\hline  
\end{tabular}  
\caption{Participants' major field (n = 32)}  
\label{table:major}
\end{table}  
\begin{table}[h]  
\centering  
\begin{tabular}{|c|c|c|c|}  
\hline  
 \textbf{Rarely} & \textbf{Sometimes} & \textbf{Often} & \textbf{Always} \\
 \hline 
 2 & 9 & 18 & 3 \\
\hline  
\end{tabular}  
\caption{Participants' reported prior usage of generative AI tools (n = 32)}  
\label{table:prior-exposure}
\end{table}  

\subsection{Participants and recruitment}
We sent a recruiting email to the students' mailing list at \IITK~inviting students to participate in a 1-hour study. The email included a sign-up form which contained questions about respondents' programming experience, programming courses taken, programming languages known, and generative AI usage. We received 71 responses before closing replies to the form.
We then invited 32 CS2-eligible respondents: 31 were undergraduate students who had taken CS1 at \IITK, and 1 was a non-CS graduate student (details of participants' majors are in Table \ref{table:major})


All participants were familiar with programming, and had prior experience using generative AI tools (Table \ref{table:prior-exposure}), but did not know JavaScript. 



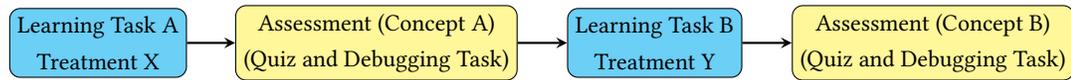
\begin{figure}[H]
\centering
\begin{tikzpicture}[node distance=3.7cm] 
\node (info1) [block, fill=cyan!50] {Learning Task A \\ Treatment X};  
\node (assessment1) [block, fill=yellow!50, right of=info1] {Assessment (Concept A)  \\ (Quiz and Debugging Task)};  

\node (info2) [block, fill=cyan!50, right of=assessment1] {Learning Task B \\ Treatment Y};  
\node (assessment2) [block, fill=yellow!50, right of=info2] {Assessment (Concept B)  \\ (Quiz and Debugging Task)};  
  
\draw [arrow] (info1) -- (assessment1);  
\draw [arrow] (assessment1) -- (info2);  
\draw [arrow] (info2) -- (assessment2);  
\end{tikzpicture} 
\caption{Study Protocol (A/B refer to Currying or IIFE; X/Y refer to AI or \NoAI) }
\label{fig1}
\end{figure}


\subsection{Study Protocol}
Each study session started with a briefing and informed consent. Participants were then given 15 minutes to learn about the concept (with or without AI) followed by an assessment. They then moved on to the learning and assessment for the second task and treatment. Figure \ref{fig1} shows the procedure of each session. For example, a participant would first learn about currying with AI and then learn about IIFE without AI.

In earlier stages of our study protocol, participants were given over 30 minutes to work on one task. However, we observed in pilot studies that given this much time, participants tended to feel frustrated and fatigued if they were unable to make progress. We thus shifted to a 10 minute limit for learning time before extending it to 15 minutes. In subsequent pilot studies, we observed that 15 minutes was sufficient for participants to gather enough information about the concept to solve our assessments. 

Participants were asked to think aloud during the learning session. If they forgot, they were prompted to do so during the session. After the learning task, the researcher conducted a retrospective interview. For the assessment, participants took a conceptual-understanding quiz followed by a 10-minute debugging task. Participants were informed about assessments during the recruitment and pre-study briefing. At the end of the debugging task, participants were asked questions about their fixes in the debugging task to determine if their task completion (if any) was accompanied by an understanding of the program, the bug, and its fix, thus indicating knowledge of the application of the concept. Participants received a cash compensation of \amount~for their participation.

\subsection{Tasks and Assessments}
\subsubsection{Tasks.} \label{learnobj} Participants were tasked with gathering information about two programming concepts associated with functional programming. All the pre-requisites for understanding these concepts (such as functions and variable scopes) were part of a mandatory CS1 course (that used C as the language of instruction) that all participants had taken. For each task, participants were provided with the following learning objectives to guide their learning: \textit{"You should be able to answer the following questions through your learning: What is <concept> in programming? Why is <concept> used in programming? How is <concept> implemented in programming languages?"
}. This was to prevent participants from being derailed by the historical or mathematical aspects of the concepts, as was the case in pilot studies. Since JavaScript was the language used in our study, we \textit{"recommended"} that participants \textit{"use JavaScript as the medium to understand the concepts"}, but did not mandate it, as we observed in pilot studies that some students preferred to learn about the concept using programming languages they knew before understanding its implementation in JavaScript. Our pilot studies also confirmed that ChatGPT consistently generated technically correct examples and explanations of both concepts.

\subsubsection{Assessment.} We evaluated participants' learning of each task using a two-part assessment. 
The first part assessed conceptual understanding via a 5-question quiz, comprising of 4 true/false questions related to the syntax and purpose of the concept and 1 multiple choice question about the usefulness of the concept. All questions also had a "don't know" option. Participants were instructed before the quiz to avoid guessing and choose this option if they did not know the answer to a particular question.  
The second part of the assessment involved participants debugging and fixing a semantic error in a JavaScript program to work according to the stated specification.  The code only contained rudimentary JavaScript structures and did not contain JavaScript-specific idioms. A patch could contain up to five lines of code. We set a 10-minute time limit for this task based on pilot studies. The quiz and debugging assessments are described in section \ref{quizA} and \ref{debugB} of the appendix respectively.

\subsection{Data Collection and Analysis} \label{scoring}

We collected the following artifacts from participants:
\begin{enumerate}
    \item Screen actions via video and audio of the study session
    \item LLM prompts and web queries written during the learning task and their timestamps
    \item Number of unique websites visited and videos watched during the web treatment of the learning task
    \item Responses to the quiz and debugging assessment
\end{enumerate}
We scored participants' responses to the two assessments for further analysis. 

For the quiz, each true/false question was awarded +1 or -1 depending on the correct / incorrect answer, while the multiple choice question (which contained multiple correct answers) was scored as +1, -1 based on all correct and incorrect answers respectively. A "don't know" response for any question received a score of 0.  
Thus, the possible quiz scores ranged from -8 to +8, with incorrect understanding carrying a greater penalty than "don't know". 

For the debugging task, we assigned three possible scores: 1) \textit{score=0}, if the final code failed to produce the desired output, 2) \textit{score=1} if the final code produced the desired output and the participant was able to correctly describe their bug fix, and 3) \textit{score=0.5}, if the code produced the desired output but the participant was unable to explain their solution. We refer to these outcomes as \textit{Incorrect}, \textit{Correct} and \textit{Correct (LoU)} respectively, where \textit{LoU} stands for lack of understanding.

\section{Results}

\subsection{RQ1: What, if any, are the differences in information-seeking strategies between the web and AI treatments?}
To answer this question, we investigated the information gathering behaviors of participants based on their screen actions during the learning session as well as their prompting/querying strategies across the two treatments. The following subsections lay out how we qualitatively coded the data and the results of this analysis.

\begin{figure}[htbp]
    \centering
    \includegraphics[width=1\textwidth]{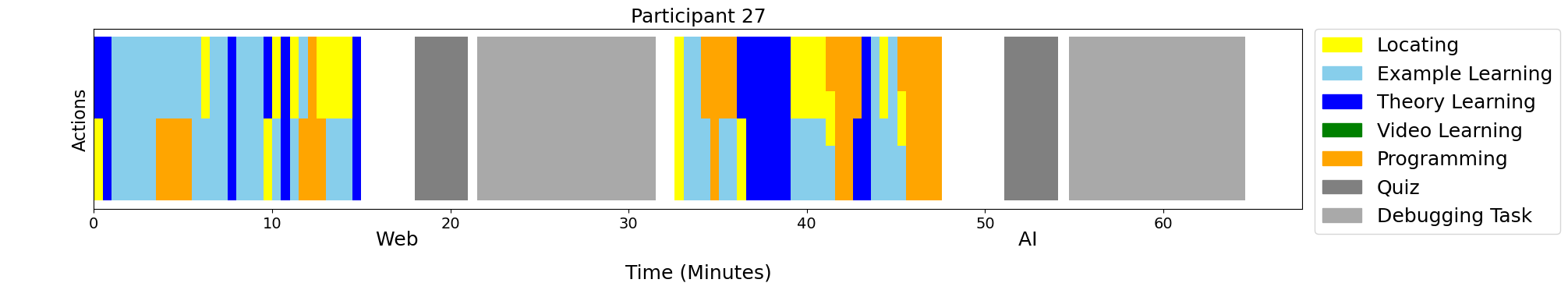}
    \caption{Example application of activity analysis codebook (vertical divisions in a segment imply multiple codes applied)}
    \label{fig:qual}
\end{figure}

\begin{table}

\begin{tabular}{|p{2cm}|p{2.4cm}|p{5.1cm}|p{4.2cm}|}
\hline
\multicolumn{1}{|c|}{\textbf{Category}} & \multicolumn{1}{|c|}{\textbf{Code}} & \multicolumn{1}{|c|}{\textbf{Definition}} & \multicolumn{1}{|c|}{\textbf{Example}} \\ [1.2ex]
\hline
\multirow{2}{*}{\textbf{Programming}}  & Coding & Participant uses the VSCode IDE while learning &  Participant writes and/or executes code on VSCode\\ [1.2ex]
\hline
\multirow{3}{*}{\textbf{Video}} & Locating & Participant looks for a video to watch & Searching "javascript currying tutorial" on YouTube \\ 
\textbf{Learning} & Watching & Participant learns by watching a video & Watching a tutorial explaining IIFE implementation \\ [1.2ex]
\hline
\multirow{6}{*}{\textbf{Learning}} & Locating & Searching for information to learn from & Scrolling through search results for a search query; Writing a prompt to ChatGPT \\
& Learning: Theory & Learning through abstract theory & Reading the definition and advantages of currying \\
& Learning: Example & Learning through information related to examples & Reading the description of an example of an IIFE \\
\hline
\end{tabular}
\caption{Activity Analysis Codebook}
\label{tab:coding-scheme}
\end{table} 

\begin{table}
\begin{tabular}{|c|c|c|c|c|}  
\hline  
\multirow{2}{*}{\textbf{Activity}} & \multirow{2}{*}{\textbf{Treatment}} & \multicolumn{3}{c|}{\textbf{Time Spent (\%)}} \\
\cline{3-5}
  &  & \textbf{Minimum} & \textbf{Maximum} & \textbf{Mean} \\
\hline  
\multirow{2}{*}{Locating} & AI & 7\% & 40\% & 26\% \\
\cline{2-5}
& \NoAI & 7\% & 38\% & 22\%\\
\hline     
Learning & AI & 16\% & 80\% & 51\% \\
\cline{2-5}
(Example) & \NoAI & 3\% & 73\% & 40\%\\
\hline  
Learning  & AI & 0\% & 40\% & 15\% \\
\cline{2-5}
(Theory) & \NoAI & 2\% & 52\% & 17\%\\
\hline  
\multirow{2}{*}{Using IDE} & AI & 0\% & 37\% & 8\% \\
\cline{2-5}
& \NoAI & 0\% & 39\% & 10\%\\
\hline
Watching Videos  & \NoAI & 0\% & 56\% & 11\% \\
\hline
\end{tabular}  
\caption{Proportion of time spent by participants across AI and \NoAI treatments}  
\label{table:actions}
\end{table}

\subsubsection{Activity Analysis}
To analyse the time participants spent on different information gathering activities, we created a codebook (Table~\ref{tab:coding-scheme}) to describe participants' screen actions. Our codes consist of programming, locating information sources, and information learning \cite{followups}, where learning was divided into theoretical learning and learning by example \cite{brandt2009twostudies}. Actions were coded in 30-second segments, with one or more codes per segment. Figure \ref{fig:qual} shows an example of the application of this coding scheme. To validate this codebook, two authors coded 4 out of 32 (>10\%) videos independently. The interrater reliability was 87\% on the Jaccard similarity index. We use the Jaccard index as the codes were not mutually exclusive and multiple code assignments per segment were allowed.

The distribution of the time spent by participants on various activities is presented in Table \ref{table:actions}. In the \NoAI treatment, the data for locating information sources includes the data for locating websites (20.7\%) as well as videos (1.4\%). As the table shows, in both conditions, participants showed a strong tendency to learn by example. P21 eloquently described why they used examples to understand concepts: \textit{``With examples, I can try to implement them and by creating variations I can understand the concept better. They are easier to comprehend... not abstract"}. Others cited that examples were helpful to understand the applications and usefulness of a concept: \textit{"It wasn't very obvious why currying was useful, but through the examples it is very clear"} (P25).

Likewise, in the \NoAI treatment, 25\% of all 32 participants watched videos, finding that videos were more accessible than websites and made concept understanding easier. For example, P22 said: 
\textit{"I felt like these websites are directly jumping to the output... I am not familiar with the syntax so it would be good if I can see the flow, and a YouTube video can help with that"}. P28 mentioned that videos can provide a ``demonstration'' of concepts and P20 mentioned that videos felt more like ``human interactions'' than websites. 

Nearly two-thirds of participants (23 out of 32) chose to try examples in the VSCode IDE while learning, across the AI and \NoAI conditions. However, this amounted to only about 10\% of their learning time on average (Table \ref{table:actions}). This can be attributed to the fact that many participants only used the IDE to run code from external sources and verify its expected output, and did not tinker with them. 7 participants mentioned that they used the IDE only to verify ChatGPT-generated code and 5 mentioned the same for websites.

\vspace{0.2cm}
\begin{mdframed}[linewidth=0.5pt, linecolor=gray!50, backgroundcolor=gray!10]

\textbf{Observations:}
\justifying 
\noindent
Participants showed a tendency to seek examples across both treatments. They did not spend much time programming during the learning task, often using the IDE only to verify external code. Some participants preferred to watch videos as a more dynamic alternative to websites.
\end{mdframed}

\begin{table}
    

\begin{tabular}{|p{2.1cm}|p{7.2cm}|p{5.2cm}|l|l|}
\hline
\multicolumn{1}{|c|}{\textbf{Code}} & \multicolumn{1}{|c|}{\textbf{Definition}} & \multicolumn{1}{|c|}{\textbf{Example}} \\
\hline
\multicolumn{3}{|l|}{\textbf{Query Origin}} \\
\hline
Copied & Queries directly pasted from external sources without modification  & Copying "What is currying in programming?" from the learning objectives \\
Edited & Queries edited or written completely by participants & Modifying the original query to "What is currying in javascript?" \\
\hline
\multicolumn{3}{|l|}{\textbf{Follow-Up}} \\
\hline
Base & Queries starting a new line of thought independent of previous queries or learned information & What is currying in javascript? \\
 Derived & Queries associated with a previous query or learned information that continue exploring an existing line of thought, such as follow-ups, modifications, or clarifications & Explain this example in more detail \\
\hline
\multicolumn{3}{|l|}{\textbf{Query Phrasing}} \\
\hline
Evidence-Based & Queries that ask about the features/description/definition of a concept/idea/object/event & What is currying in programming?\\  
Reason &  You want to find out reasons of/for something. & Why are IIFEs used? \\  
Comparison & Queries that ask to compare/contrast two or more things, understand their differences/similarities. & How are curried functions better than normal functions? \\  
Instruction & Queries to understand the procedure/method of doing/achieving something. & How do I write an IIFE in javascript? \\  
Debate &  Queries that debate a hypothetical
 question (is someone right or wrong,
 is some event perceived positively or
 negatively?). & Do you really think currying can be useful?\\  
Experience & Queries that ask for advice
 or recommendations
 on a particular topic. &  Do you recommend using currying? \\  
Keyword-Based & Queries not phrased as a question that do not solicit information & Currying in javascript \\
Explanation & Queries not phrased as a question that solicit information & Please explain how currying helps with function composition \\
\hline
\end{tabular}

\caption{Query Classification Codebook}
\label{tab:query}
\end{table}

\begin{table}
\centering  
\begin{tabular}{|l|l|l|l|}  
\hline  
\textbf{Property} & \textbf{Web queries in \NoAI} & \textbf{Prompts in AI} \\  
\hline  
Total & 148 & 237 \\  
\hline  
\multicolumn{3}{|l|}{\textbf{Follow-up}} \\
\hline  
Base  & 67\% & 53\% \\  
Derived & 33\% & 47\% \\ 
\hline  
\multicolumn{3}{|l|}{\textbf{Query Origin}} \\  
\hline  
Copied & 20\% & 13\% \\  
Edited & 80\% & 87\% \\  
\hline  
\multicolumn{3}{|l|}{\textbf{Query Phrasing}} \\ 
\hline  
Evidence-Based  & 29\% & 39\%\\  
Reason  & 13\% & 12\% \\  
Instruction & 5\% & 4\% \\ 
Comparison & <1\% & 3\% \\   
Debate & <1\% & 0  \\  
Experience & 0\% & <1\% \\  
Keyword-Based & 42\% & 4\% \\  
Explanation & 11\% & 39\% \\  
\hline  
\end{tabular}  
\caption{Prompt and Query Analysis across participants}  
\label{table:1}   
\end{table}

\subsubsection{Prompt and Query Analysis}
Participants wrote a total of 237 prompts to ChatGPT and a total of 148 search queries (Google + YouTube) in the \NoAI treatment. To understand the nature of these prompts and queries, we categorized them across three dimensions: follow-up, origin, and phrasing (Table~\ref{tab:query}). As web searches often lead to follow-up queries \cite{followups}, this analysis helps us examine if and how LLM prompts lead to follow-ups. To understand phrasing, we adapted the `Expression' dimension of the TELer taxonomy of prompts by Karmaker et al.\cite{karmaker-santu-feng-2023-teler}, dividing prompts and queries into question-style and instruction-style. The taxonomy for question-style phrasing has been adapted from Bolotova et al.'s \cite{taxonomy} taxonomy for questions. For instruction-style queries, we introduce two 
subcategories - \textit{Keyword-Based} and \textit{Explanation}, described in Table~\ref{tab:query}. Based on coding the prompts and queries of 4 participants, the interrater reliability was 89\%, 83\% and 81\% on Cohen's Kappa~\cite{kappa} for the dimensions of origin, follow-up, and phrasing respectively.
Table~\ref{table:1} lists these code occurrences in the web and ChatGPT conditions. 

Participants began their learning episodes using the learning objectives provided to them. Nearly half of all participants (15 out 32) used the text of the learning objectives (described in Section \ref{learnobj}) directly as the starting point for prompting in the AI condition, in contrast to only 5 out of 32 in web search (\NoAI). Most (18 out of 32) participants used keywords based on the learning objectives (queries phrased as \textit{Keyword-Based}) as the starting point for their web information seeking. 

As participants progressed in their learning, querying differences surfaced again. We analysed the distribution of follow-ups across treatments by applying the chi-square test of independence, as we have over 30 subjects and a sufficiently large sample size of queries and prompts ($N = 385$). This test indicated that LLM prompts were more likely to be follow-ups than web search queries ( $\chi^2 (1, N = 385) = 6.93, p = .008$). Table~\ref{table:1} also suggests that a higher percentage of web queries were copied directly, from sources such as prior websites they had visited. These differences in information-seeking behaviors could be considered adaptations to the environment \cite{pirolli1995}. For example, ChatGPT offered byte-sized answers to specific questions and afforded easy follow-up questions in a more chatty manner, whereas web searches offered a longer list of diverse, potentially useful \textit{sources} that participants had to sift through and choose to learn from \cite{sellen2002knowledge}. As P22 noted for learning with the web, \textit{``You don't know how reliable one source is, and what new or what different you might find from another source''.}

Drilling further into the differences in learning behaviors between the AI and \NoAI environments, we evaluated the effects of learning treatment and phrasing type (Table~\ref{tab:query}; query phrasing) on the number of queries/prompts issued by participants. We modeled this count using a Generalized Linear Model (GLM) with a Poisson distribution and a log link function (dispersion parameter = 1.19). We found a significant interaction effect between learning treatment and query phrasing ($\chi^2 (4, N = 320    ) = 115.11, p = < 0.001$) and a significant main effect for learning treatment ($\chi^2 (1, N = 320) = 27.08, p = < 0.001$) using the chi-square test ($N$ indicates the total number of observations, each one corresponding to one learning treatment and query type for one participant). This main effect indicates that participants used a significantly higher number of prompts in the AI treatment than queries in the \NoAI treatment, reiterating the more interactive nature of LLM-based learning.


The significant interaction effect suggests that specific learning environments promote or discourage certain ways of phrasing queries more than others. 
Table \ref{table:1} hints at these instances of query phrasing differences across the two treatments. For example, participants used more explanation-soliciting (\textit{Explanation}) queries in the AI treatment, while in the \NoAI treatment they tended to use keyword-based queries (\textit{Keyword-Based}), suggesting a preference for more generic than specific information seeking in the web. This also reflects in the use of the word ``example'' in only  7\% of web search queries as opposed to in 20\% of LLM prompts, though participants spent high amounts of time learning by examples in both treatments (Table ~\ref{table:actions}). Similarly, 57\% of LLM prompts were phrased as questions, as compared to 47\% in search queries. During a retrospective interview, P31 explained their reason for phrasing queries as questions to ChatGPT but not to Google: \textit{ ``Google doesn't behave like ChatGPT.... instead of saying `what is this', it's better to just write the key terms for Google''}.

\begin{figure}[htbp]
    \centering
    \includegraphics[width=\textwidth]{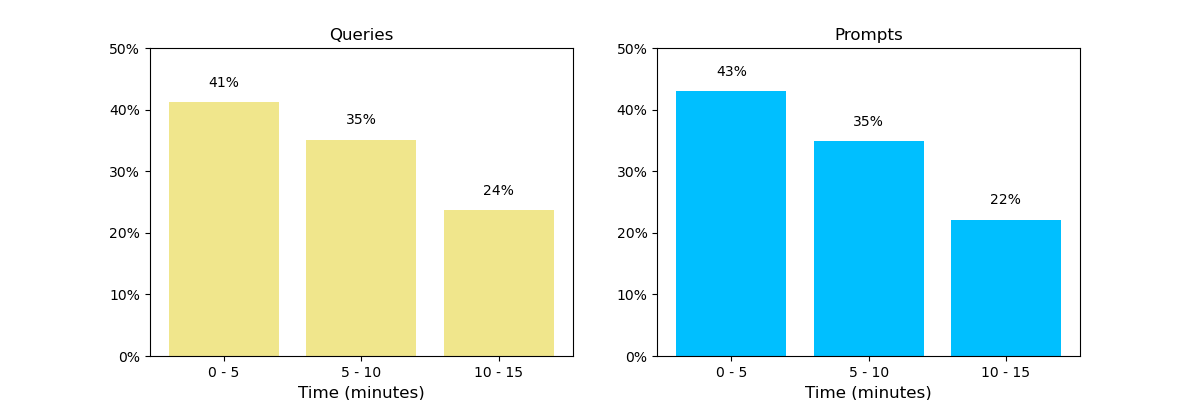}
    \caption{Percentage of prompts and queries over time}
    \label{fig:query_time}
\end{figure}

Across both treatments, participants showed a tendency to ask more queries during the start of the learning process. Figure \ref{fig:query_time} shows the distribution of the number of queries and prompts entered by participants over the learning session. 



\vspace{0.2cm}
\begin{mdframed}[linewidth=0.5pt, linecolor=gray!50, backgroundcolor=gray!10]

\textbf{Observations:}
\justifying 
\noindent
Participants used more follow-up queries when using LLMs, and phrased their queries differently across learning conditions, showing a tendency to use keyword-based queries for search engines and to explicitly ask for information in their queries to LLMs.
\end{mdframed}

\subsection{RQ2: How effective are the information-seeking strategies in the web and AI treatments in terms of conceptual understanding and practical application? }

To understand the effectiveness of participants' information-seeking strategies, we quantitatively analysed participants' solutions to the quiz and debugging task based on the scoring metric outlined in Section \ref{scoring}. 

\subsubsection{Debugging Task Performance.} 
\begin{figure}[htbp]
    \centering
    \includegraphics[width=0.6\textwidth]{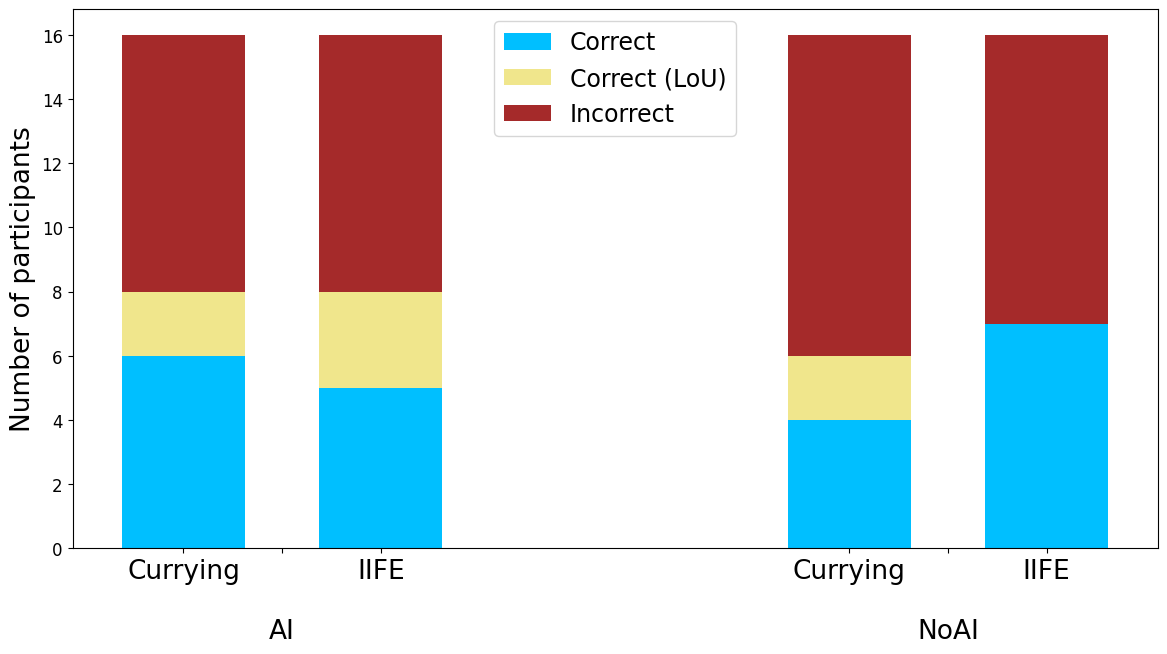}
    \caption{Debugging Task Performance (\textit{LoU} refers to lack of understanding)}
    \label{fig:debug}
    \vspace{0.3cm}
    \centering
    \includegraphics[width=0.9\textwidth]{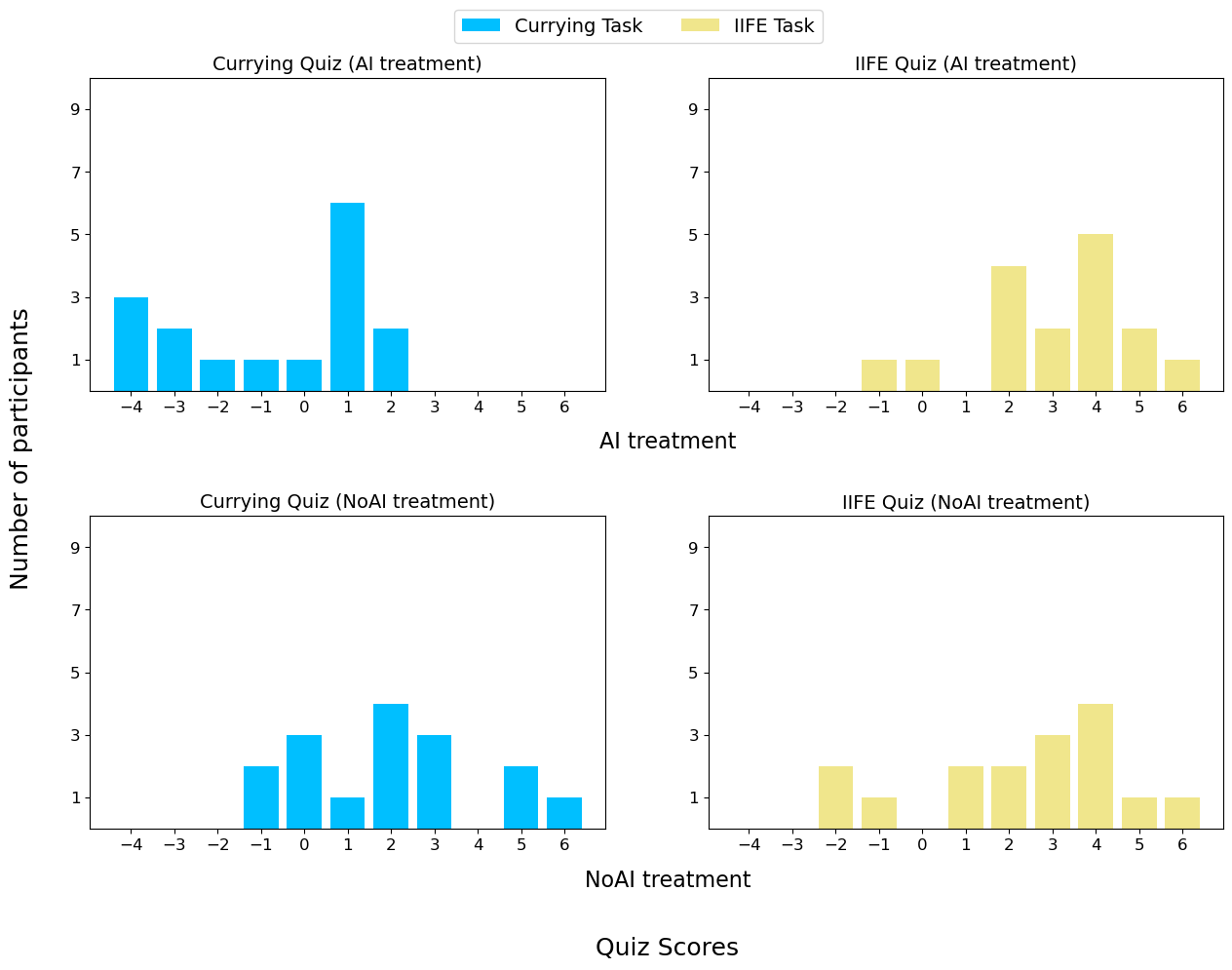}
    \caption{Conceptual understanding Quiz Scores }
    \label{fig:quiz}
\end{figure}

Figure \ref{fig:debug} gives an overview of the results of the debugging task across tasks and treatments. We use the paired t-test to analyse these results as our data is paired and we have $N= 32$ per distribution owing to our within-subject study design. 
We did not find significant differences across the dimensions of learning task ($t(31) = 1.79, p = .08$) or learning treatment ($t(31) = 0.57, p = .57$). To compare debugging performance for a particular learning task or with a particular learning treatment, we have a total of 32 samples, and thus apply the Shapiro-Wilk normality test on both distributions to decide whether to apply parametric or non-parametric tests. We also use tests for independent data as participants were given a different task for each treatment. We did not find any significant differences on across the dimensions of learning treatment when considering the debugging task performance for currying ($U = 143.0, p = .53$) and  IIFE ($U = 124.0, p = .88$). We use the Mann-Whitney U-test as the Shapiro-Wilk test indicated that the debugging scores with the \NoAI treatment for the currying task ($W= 0.64, p <0.001$) and the IIFE task ($W= 0.74, p <0.001$) were not normally distributed. 

\subsubsection{Quiz Performance}
Figure \ref{fig:quiz} presents the distribution of quiz scores for each concept and learning treatment, scored on a scale of [-8, +8]. We found that the quiz scores for currying ($M = 0.66, SD = 2.52$) were significantly lower  than IIFE ($M = 2.69, SD = 2.10$), based on the paired t-test ($t(31) = -4.21, p < 0.001$). The effect size, measured by Cohen’s d, was $d = -0.86$, indicating a large effect.
It also took participants significantly more time to complete the currying quiz compared to IIFE (163.1 vs 97.21 seconds, on average) as per the paired t-test ($t(31) = 5.17, p < 0.001$). Across all participants and questions in the quiz, participants chose the \textit{"don't know"} option 31 times (15.9\% of all options chosen) in the quiz for currying, and 29 times (13.5\% of all options chosen) in the quiz for IIFE. This may suggest that currying is a more difficult topic for participants to learn in the study condition than IIFE in general, or that the post-study quiz in currying task was harder than in IIFE. As a result, we compare the \NoAI and AI conditions for each task. 
\paragraph{AI vs. \NoAI in Currying} Our sample size here is small ($N = 16$ per treatment) and the currying quiz scores for the AI treatment were non-normal [Shapiro-Wilk ($W = 0.84, p = 0.01$)]. Therefore, we conducted a Mann-Whitney U-test and found statistical differences between the mean quiz scores in the AI ($M = -0.69, SD = 2.2$) and \NoAI ($M = 2.0, SD = 2.06$) treatments ($-0.69 \text{ vs. } 2.0, p=.005$). The effect size was large ($r = 0.50$).  This difference is despite the fact that participants' information-seeking activities for currying were similar across both treatments (Table \ref{table:curryactions}). Participants in the \NoAI treatment were more likely to use \textit{Keyword-Based} queries (51\% as compared to the overall average of 42\%) for this task (Table \ref{table:curryqueries}).
\paragraph{AI vs. \NoAI in IIFE} In contrast, we did not find any significant difference in the IIFE quiz scores between the AI and \NoAI conditions based on the t-test for independent samples ($t(30) = 0.99, p = 0.33$). We use this test since Shapiro-Wilk tests suggested normal distributions of IIFE quiz scores in AI ($W = 0.92, p =.20$) and \NoAI ($W = 0.93, p = .28$) treatments. 

\vspace{0.2cm}
\begin{mdframed}[linewidth=0.5pt, linecolor=gray!50, backgroundcolor=gray!10]

\textbf{Observations:}
\justifying 
\noindent
Participants learning with AI performed significantly worse than those learning through web search in the theoretical quiz for currying.
\end{mdframed}

\section{Threats to validity}

\noindent{\textbf{Internal.}}
We randomized the order of tasks and treatments to minimize learning effects. We had two tasks with different difficulty levels and this stimulated diversity in information gathering and learning tasks.

\noindent{\textbf{Construct.}}
Our assessments involved quizzes and debugging tasks, which are insufficient to evaluate long-duration participant learning and retention. However, we validated our quiz questions and the debugging task in pilot studies to evaluate participants' knowledge on key concepts in the learning goals. 

\noindent{\textbf{External.}}
Our results are based on a small sample size relative to the general student population. \IITK~has a female:male ratio of 20\%, and despite our best efforts we could not recruit a gender balanced sample. Only ~20\% of our participants were female. Replication of this study with a gender-balanced sample is needed to generalize our results to other environments. There is also a need for further studies to generalize our findings to other kinds of learning tasks and situations (e.g., concepts in computer architecture or usability).

\section{Discussion and Concluding Remarks}


In this paper, we provide a comparative study on how students used AI (ChatGPT) and traditional web resources (Google search and YouTube) to gather information on new topics in programming in a self-guided manner.
As ChatGPT and similar AI tools gain popularity, examining them as learning tools in the context of traditional online learning methods helps us understand their efficacy and applicability in different learning scenarios.

\paragraph{Perform Fast and Learn Slow}
ChatGPT and other LLMs can improve students' productivity in performing programming tasks~\cite{kazemitabaar2023novices}, but their effectiveness in helping students learn and understand new concepts is unclear. In our study, while AI helped participants slightly more than web search in successfully debugging programs, a notable number of participants in the AI condition lacked understanding of their proposed fixes (Figure~\ref{fig:debug}). Similarly, participants in the AI treatment performed poorly in the quiz for currying, the more difficult task, as compared to participants in the \NoAI treatment, suggesting that the nature of productivity may be at odds with learning. While productivity emphasizes on the speed in completion of tasks, successful learning can be slow and tedious, and ChatGPT might not be the best tool for learning, especially in the absence of sufficient metacognitive abilities \cite{cs1slow, prather2024widening}. 

\paragraph{New Tools, New Challenges}
From Table \ref{table:actions}, we can observe that the average time participants spent in the AI and \NoAI treatment towards programming (8\% vs. 10\%), locating information (26\% vs. 22\%), and learning information (66\% vs. 68\%) are similar. Despite the fact that web search often requires users to navigate through multiple information sources \cite{sellen2002knowledge} and that ChatGPT can respond directly to user queries to generate explanations and examples, participants spent more time locating information with ChatGPT. This could be due to information-seeking tasks specific to the AI treatment such as thinking of and writing prompts to express intent \cite{nguyen2024beginning} and scrolling through the chat history to find information from previous responses. Analysing which learning scenarios might be best suited to address the unique challenges of these different information-seeking setups can help define the role of LLMs as learning tools for computing education.

\paragraph{Importance of Example-based learning}
Learning theories have emphasized the importance of examples in learning. Examples can provide learning opportunities for learners to imitate experts, discern analogies and contrast among different problems. In our study, participants across treatments showed a tendency to learn through examples -- participants in the AI treatment spent 51\% of their time learning by example, while those in the \NoAI treatment spent 40\% of the time learning by example and an additional 10\% learning through videos. This emphasizes the importance of incorporating high-quality examples in pedagogy. 

\paragraph{Need for Holistic LLM Responses when Learning}
Participants in our study tended to write keyword-based queries when using search engines, which led them to websites containing human-curated content about programming concepts.
While using LLMs, participants wrote more questions and explicitly asked for information. Since LLMs reply directly to user prompts, their responses might only cover the portion of information about the concept that is asked for in the prompt. Our results suggest that while using LLMs, students are more likely to ask follow-ups and thus dive deeper into areas they have already explored previously. For the more difficult task, 51\% of participants' queries in the \NoAI treatment were keyword-based, as compared to less than 1\% for participants in the AI treatment (Table \ref{table:curryqueries}). Using more specific prompts might lead learners down a rabbit hole, preventing them from gaining a more complete understanding of the concept, potentially contributing to participants' poorer quiz performance in the AI treatment for this task.  
Future research can explore how LLMs can understand student intent and needs to provide more holistic responses rather than generating direct responses to prompts in exploratory learning scenarios. One approach would be to draw inspiration directly from web searches that afford learning from multiple perspectives by utilizing multiple LLM responses. 

\paragraph{Text is NOT enough!}
Modern LLMs generate high-quality textual responses to user prompts. They can provide examples as well as high-quality explanations of these examples \cite{leinonen2023comparing}.
However, participants in the \NoAI treatment also expressed interest in non-textual learning resources, notably videos. Some participants chose to watch videos as they are more dynamic than text-based responses, and allow them to better understand the process behind using the concept. Others cited them to be more human. These preferences are consistent with prior work by Moghavvemi et al. \cite{MOGHAVVEMI201837} on the effectiveness and popularity of YouTube as a tool for information seeking. Thus, when choosing generative AI tools as learning resources in educational settings, it is imperative to choose ones that allow for heterogenous content types for different learning preferences and styles. For example,  \cite{PerplexityAI} can augment its generated textual content with links to video resources. However, there are opportunities for research and development in multimodal language models that can help in creating more effective learning content (e.g., videos, interactive visualizations) and human-like experiences for diverse subjects and learners \cite{lee2023multimodalityaieducationartificial}.

\appendix
\section*{Appendix}  

\section{Conceptual Understanding Quiz}  \label{quizA}
Tables \ref{tab:currying_questionnaire} and \ref{tab:iife_questionnaire} display the questions participants answered in the theory assessment for the Currying and IIFE learning task respectively. The correct options are filled in ($\blacksquare$).
\begin{table}[H]
\centering
\renewcommand{\arraystretch}{0.77}
\begin{tabular}{|p{0.8\textwidth}|p{0.15\textwidth}|}
\hline
\textbf{Question} & \textbf{Options} \\
\hline
Passing arguments (for full application of the function) to a curried function and a non-curried function is done with the same syntax. & 
\begin{tabular}[c]{@{}l@{}}
$\square$ Don't Know\\
$\square$ Yes\\
$\blacksquare$ No
\end{tabular} \\
\hline
The curried version of a function that takes one argument is the same as the non-curried version. & 
\begin{tabular}[c]{@{}l@{}}
$\square$ Don't Know\\
$\blacksquare$ Yes\\
$\square$ No
\end{tabular} \\
\hline
Curried functions always return values (variables, constants, string, etc.) when they are called with any number of arguments. & 
\begin{tabular}[c]{@{}l@{}}
$\square$ Don't Know\\
$\square$ Yes\\
$\blacksquare$ No
\end{tabular} \\
\hline
Every nested function is an example of currying. & 
\begin{tabular}[c]{@{}l@{}}
$\square$ Don't Know\\
$\square$ Yes\\
$\blacksquare$ No
\end{tabular} \\
\hline
\multicolumn{2}{|p{\dimexpr\textwidth-2\tabcolsep\relax}|}{\textbf{Which of the following are the advantages of Currying a function?} (Select one or more)} \\
\hline
\multicolumn{2}{|p{\dimexpr\textwidth-2\tabcolsep\relax}|}{
\begin{tabular}[c]{@{}l@{}}
$\square$ Don't Know\\
 $\blacksquare$ We can create functions with partial functionality of the original curried function.\\
$\square$ The curried version of a function can return the final value with a fewer number of arguments.\\
 $\blacksquare$ We can avoid passing the same variable many times to a function by currying.\\
$\square$ We can add extra functionality by currying that is not present in the non-curried function.
\end{tabular}
} \\
\hline
\end{tabular}
\caption{Currying Conceptual understanding Quiz}
\label{tab:currying_questionnaire}
\end{table}


\begin{table}[H]
\centering
\renewcommand{\arraystretch}{0.77}
\begin{tabular}{|p{0.8\textwidth}|p{0.15\textwidth}|}
\hline
\textbf{Question} & \textbf{Options} \\
\hline
The variables created within the body of an IIFE can be accessed outside the IIFE. & 
\begin{tabular}[c]{@{}l@{}}
$\square$ Don't Know\\
$\square$ Yes\\
 $\blacksquare$ No
\end{tabular} \\
\hline
IIFEs can never have a name. & 
\begin{tabular}[c]{@{}l@{}}
$\square$ Don't Know\\
$\square$ Yes\\
 $\blacksquare$ No
\end{tabular} \\
\hline
IIFEs can be called later in code after they are defined. & 
\begin{tabular}[c]{@{}l@{}}
$\square$ Don't Know\\
$\square$ Yes\\
 $\blacksquare$ No
\end{tabular} \\
\hline
We can pass arguments to IIFEs. & 
\begin{tabular}[c]{@{}l@{}}
$\square$ Don't Know\\
 $\blacksquare$ Yes\\
$\square$ No
\end{tabular} \\
\hline
\multicolumn{2}{|p{\dimexpr\textwidth-2\tabcolsep\relax}|}{\textbf{Which of the following are the advantages of IIFEs?} (Select one or more)} \\
\hline
\multicolumn{2}{|p{\dimexpr\textwidth-2\tabcolsep\relax}|}{
\begin{tabular}[c]{@{}l@{}}
$\square$ Don't Know\\
 $\blacksquare$ It allows us to reduce the number of global variables\\
$\square$ It is always invoked immediately whenever the compiler sees it.\\
 $\blacksquare$ It allows us to isolate a particular part of code functionality.\\
$\square$ It can have extra functionality that cannot be done with traditional functions.
\end{tabular}
} \\
\hline
\end{tabular}
\caption{IIFE Conceptual understanding Quiz}
\label{tab:iife_questionnaire}
\end{table}

\section{Debugging Assessment} \label{debugB}
Figures \ref{fig:iife_ps} - \ref{fig:curry_correct} show the problem statement and relevant snippets of the expected solution for the debugging assessment for the currying and IIFE learning tasks respectively. Participants did not necessarily have to produce code identical to the expected solution, but had to reach the desired output as specified in the problem statement without hardcoding the solution. The results were manually verified by the researcher conducted the study.
\begin{figure}[H]
\begin{lstlisting}[label=lst:iife_bug, frame=single, framexleftmargin=0mm, framexrightmargin=0mm, framesep=2mm, basicstyle=\ttfamily\scriptsize, aboveskip=3pt, belowskip=3pt]
// This code uses an IIFE to print the sum of a and b. 
// Why isn't it working? Edit the code to fix it, keeping the structure the same.

var a = 20;
var b = 40;

//Start changes below here!

var result;

function adder(){
  (function(a, b) {
    var result = a + b;
  })(a, b);
}

console.log(result);
\end{lstlisting}
\caption{IIFE Debugging Task - Problem Statement}
\label{fig:iife_ps}
\end{figure}

\begin{figure}[H]
\begin{lstlisting}[ label=lst:iife_correct, frame=single, framexleftmargin=0mm, framexrightmargin=0mm, framesep=2mm, basicstyle=\ttfamily\scriptsize, aboveskip=3pt, belowskip=3pt]
....
....
//Start changes below here!

var result;

function adder(){
  (function(a, b) {
    result = a + b;
  })(a, b);
}

adder();
console.log(result);
\end{lstlisting}
\caption{IIFE Debugging Task - Expected Solution (Snippet)}
\label{fig:iife_correct}
\end{figure}

\begin{figure}[H]
\begin{lstlisting}[label=lst:curry_bug, frame=single, framexleftmargin=0mm, framexrightmargin=0mm, framesep=2mm, basicstyle=\ttfamily\scriptsize, aboveskip=3pt, belowskip=3pt]
// This code is supposed to print the values of a, b, and c once, and then print the sum on the next line.
// To clarify, the expected output is:
// a fixed! Value of a: 
// 1
// b fixed! Value of b: 
// 2
// c fixed! Value of c: 
// 3
// 6
// Currently, the code is not working. But you can use the fact that sum is a curried function to fix it!
// Edit the code to make it work as intended without changing the code up to line 30.

//Tip: Take a couple of minutes first to understand the function below

//Do not change
function sum(a) {
  console.log('a fixed! Value of a: ')
  return (b) => {
    console.log('b fixed! Value of b: ')
    return (c) => {
        console.log('c fixed! Value of c: ')
        return a + b + c
      }
  }
}
var a = 1;
var b = 2;
var c = 3;

//Start changes below here!

suma = sum(a)(b)(c)
console.log(a)
sumb = sum(a)(b)(c)
console.log(b)
sumc = sum(a)(b)(c)
console.log(c)
console.log(sumc)
\end{lstlisting}
\caption{Currying Debugging Task - Problem Statement}
\label{fig:curry_ps}
\end{figure}

\begin{figure}[H]
\begin{lstlisting}[label=lst:curry_correct, frame=single, framexleftmargin=0mm, framexrightmargin=0mm, framesep=2mm, basicstyle=\ttfamily\scriptsize, aboveskip=3pt, belowskip=3pt]
...
...
//Start changes below here!

suma = sum(a)
console.log(a)
sumb = suma(b)
console.log(b)
sumc = sumb(c)
console.log(c)
console.log(sumc)
\end{lstlisting}
\caption{Currying Debugging Task - Expected Solution (Snippet)}
\label{fig:curry_correct}
\end{figure}


\section{Currying Task Qualitative Data}
Table \ref{table:curryactions} and Table \ref{table:curryqueries} show the data for participant activities and prompts/queries respectively for the currying learning task.

\begin{table}[H]
\begin{tabular}{|c|c|c|}  
\hline  
\multirow{1}{*}{\textbf{Activity}} & \multirow{1}{*}{\textbf{Treatment}} & \multicolumn{1}{c|}{\textbf{Average Time Spent (\%)}} \\
\hline  
\multirow{2}{*}{Locating} & AI & 24\% \\
& \NoAI & 24\% \\
\hline     
Learning & AI & 56\% \\
(Example) & \NoAI & 41\%\\
\hline  
Learning  & AI & 12\% \\
(Theory) & \NoAI & 13\%\\
\hline  
\multirow{2}{*}{Using IDE} & AI & 7\% \\
& \NoAI &  6\%\\
\hline
Watching Videos  & \NoAI & 16\% \\
\hline
\end{tabular}  
\caption{Proportion of time spent by participants in the currying learning task across AI and \NoAI treatments}  
\label{table:curryactions}
\end{table}

\begin{table}[H]
\centering  
\begin{tabular}{|l|l|l|l|}  
\hline  
\textbf{Property} & \textbf{Web queries in \NoAI} & \textbf{Prompts in AI} \\  
\hline  
Total & 69 & 108 \\  
\hline  
\multicolumn{3}{|l|}{\textbf{Follow-up}} \\
\hline  
Base  & 72\% & 56\% \\  
Derived & 28\% & 44\% \\ 
\hline  
\multicolumn{3}{|l|}{\textbf{Query Origin}} \\  
\hline  
Copied & 14\% & 14\% \\  
Edited & 86\% & 86\% \\  
\hline  
\multicolumn{3}{|l|}{\textbf{Query Phrasing}} \\ 
\hline  
Evidence-Based  & 27\% & 43\%\\  
Reason  & 8\% & 12\% \\  
Instruction & 6\% & 4\% \\ 
Comparison & 1\% & 3\% \\   
Debate & 0\% & 0\%  \\  
Experience & 0\% & <1\% \\  
Keyword-Based & 51\% & <1\% \\  
Explanation & 7\% & 37\% \\  
\hline  
\end{tabular}  
\caption{Prompt and Query Analysis across participants for the currying learning task}  
\label{table:curryqueries}   
\end{table}

\bibliography{references}
\end{document}